\documentclass[aps,prc,twocolumn,superscriptaddress,groupedaddress]{revtex4} 
\usepackage{graphicx}  
\usepackage{dcolumn}   
\usepackage{bm}        
\usepackage{amssymb}   
\usepackage{amsfonts}
\usepackage{booktabs}
\usepackage{siunitx}
\usepackage{xcolor}
\usepackage{multirow}
\usepackage[normalem]{ulem}
\usepackage{amsmath}

\begin{document}

\title{Three-State Mixing as a Phenomenological Framework\\for Multiple Shape Coexistence}
\author{M.~Siciliano}
    \email{msiciliano@anl.gov}
    \affiliation{Physics Division, Argonne National Laboratory, Lemont (IL), United States.}

\begin{abstract}
\noindent
Shape coexistence represents one of the most striking manifestations of competing collective and single-particle degrees of freedom in atomic nuclei. 
While the coexistence and mixing of two configurations can be described within the well-established Two-State Mixing framework, the observation of three or more competing structures requires a more general treatment. 
In this work, we introduce a Three-State Mixing (3SM) model in which three intrinsic configurations are related to the physical states through an $SO(3)$ rotation. 
The framework establishes a direct connection between experimental observables, configuration-mixing amplitudes, and intrinsic properties, while the experimentally known excitation energies allow the corresponding effective Hamiltonian and interaction strengths to be reconstructed. 
The model is applied to the low-lying structure of $^{116}$Sn using electromagnetic matrix elements recently determined through a comprehensive Coulomb-excitation measurement. 
The analysis identifies three intrinsic configurations characterized by spherical, weakly oblate, and strongly deformed triaxial shapes, together with substantial configuration mixing among the physical $0^+$ states. 
The present formulation provides a general phenomenological framework for investigating systems in which multiple configurations coexist and strongly interact.
\end{abstract}

\maketitle


Shape coexistence is one of the clearest manifestations of competing correlations in finite quantum many-body systems. 
It arises when distinct collective configurations, associated with different intrinsic shapes, occur within a narrow excitation-energy range in the same nucleus and often interact strongly through configuration mixing~\cite{heyde2011shape, shapecoexistence2022, LEONI2024104119}. 
During the last two decades, the phenomenon has been identified throughout the nuclear chart, from light nuclei to the neutron-deficient Pb region, demonstrating that the competition between spherical shell effects and collective deformation is a universal feature of the nuclear interaction~\cite{andreyev2000, hadynska2016ca, leoni2017ni, ojala2022pb}. 
Although the coexistence of two structures is now recognized throughout the nuclear chart, the simultaneous presence and mixing of three or more low-lying configurations --referred to as \emph{multiple-shape coexistence}-- remains considerably rarer and poses a more stringent challenge to both experiment and theory. 
Only a handful of nuclei currently exhibit convincing evidence for this phenomenon, including cases in the Ni, Cd, Hg, and Pb regions~\cite{leoni2017ni, garrett2019multiple, garrett2020multiple, siciliano2020coexistence, ojala2022pb}. 
Such systems represent some of the most stringent benchmarks for microscopic descriptions of the nuclear many-body problem.

Experimentally, establishing multiple-shape coexistence is considerably more demanding than identifying the traditional two-state scenario. 
Excitation-energy systematics, enhanced electric-monopole transitions, rotational-like band structures, and selective population in transfer reactions often provide compelling indications for the coexistence of several configurations~\cite{wood1999E0, shapecoexistence2022, LEONI2024104119}. 
However, these observables alone generally do not permit an unambiguous determination of the intrinsic deformation associated with each structure. 
In contrast, high-precision Coulomb-excitation measurements provide direct access to signed electromagnetic matrix elements (MEs) and, through quadrupole sum rules, enable a model-independent determination of intrinsic quadrupole moments, triaxiality, and deformation strength~\cite{kumar1972, gosia1986}. 
Despite their importance, experimental studies reaching this level of structural information remain extremely scarce.

Recently, a high-precision multi-step Coulomb-excitation experiment on the semi-magic nucleus $^{116}$Sn produced one of the most complete sets of signed electromagnetic MEs available for a multiple-shape coexistence candidate. 
The experimental setup, data analysis, and model-independent determination of the intrinsic shapes of the three lowest $0^+$ states are presented in Ref.~\cite{Siciliano116SnArxiv}. 
Enhanced $E0$ transitions among low-lying $0^+$ states in $^{116}$Sn point to strong configuration mixing, while the markedly different population patterns observed in two-proton transfer reactions reveal distinct proton components in their wave functions~\cite{BACKLIN1981490, FIELDING1977389, BRON1979335, shapecoexistence2022, LEONI2024104119}. 
Rotational-like sequences and the characteristic excitation-energy systematics of several low-lying structures further suggest the presence of more than one intruder configuration. 

Motivated by these unique experimental constraints, the present contribution introduces a phenomenological Three-State Mixing (3SM) model that extends the traditional Two-State Mixing formalism~\cite{clement2007kr, siciliano2020coexistence} to three interacting configurations. 
The model is formulated within the $SO(3)$ rotation group and provides a transparent framework to extract intrinsic electromagnetic properties, effective interaction strengths, and unperturbed excitation energies directly from the experimental observables. 
Its application to $^{116}$Sn is presented as a representative example.

\section{Three-State Mixing Model}


The phenomenology of shape coexistence is commonly interpreted in terms of the mixing between distinct intrinsic configurations. 
Within the traditional Two-State Mixing (2SM) model~\cite{clement2007kr, siciliano2020coexistence}, the experimentally observed states are described as linear combinations of two unperturbed configurations, whose interaction gives rise to various observables, such as the measured excitation energies and electromagnetic transition strengths~\cite{Wood1992, heyde2011shape}. 

In the present work, we formulate an extension of the 2SM formalism by considering three intrinsic configurations, denoted as 
$\left\{ |J^\pi_A\rangle, \, |J^\pi_B\rangle, \, |J^\pi_C\rangle \right\}$, 
which constitute an orthonormal basis (i.e., $\langle J^\pi_i | J^\pi_j \rangle \!=\! \delta_{ij}$ with $i,j\!=\!A,B,C$). 
The experimentally observed (physical) states 
$\left\{ |J^\pi_1\rangle, \, |J^\pi_2\rangle, \, |J^\pi_3\rangle \right\}$ 
are assumed to arise from the linear combination of these intrinsic configurations. 
Since both sets of wave functions form orthonormal bases spanning the same Hilbert subspace, the transformation relating them must preserve the scalar product. 
The mixing matrix $R$ must be orthogonal (i.e., $R^{\mathrm T}R\!=\!I$ where $I$ denotes the identity matrix); furthermore, imposing $\det(R)\!=\!+1$ restricts the transformation to proper rotations and eliminates unphysical reflections of the basis states. 
Consequently, the transformation belongs to the special orthogonal group $SO(3)$. 
The physical and intrinsic bases are therefore related through
\begin{equation}
\label{eq:rotation}
\begin{pmatrix}
|J^\pi_1\rangle\\
|J^\pi_2\rangle\\
|J^\pi_3\rangle
\end{pmatrix}
=
R
\begin{pmatrix}
|J^\pi_A\rangle\\
|J^\pi_B\rangle\\
|J^\pi_C\rangle
\end{pmatrix},
\end{equation}
where
\begin{equation}
R=
\begin{pmatrix}
a_{1A} & a_{1B} & a_{1C}\\
a_{2A} & a_{2B} & a_{2C}\\
a_{3A} & a_{3B} & a_{3C}
\end{pmatrix},
\end{equation}
and the coefficients $a_{ij}$ represent the mixing amplitudes between the intrinsic and physical configurations. 
It is important to emphasize that the intrinsic configurations $|J^\pi_A\rangle$, $|J^\pi_B\rangle$, and $|J^\pi_C\rangle$ are introduced as a general orthonormal basis spanning the relevant subspace of the Hilbert space. 
No assumptions are made regarding their microscopic origin, which may differ from one nucleus to another. 
Depending on the specific case, they may correspond to normal and intruder configurations, states characterized by different equilibrium deformations, or more general shell-model configurations~\cite{PhysRevC.94.051303, Macchiavelli2017Mg, PhysRevLett.128.252501}. 
Consequently, the formalism developed below is independent of any particular microscopic interpretation and is applicable to any system exhibiting the mixing of three interacting configurations. 

Any element of $SO(3)$ may be parametrized by three independent Euler angles. 
Adopting the conventional $R_z(\alpha)R_y(\beta)R_z(\gamma)$ decomposition, the rotation matrix can be written as $R(\alpha,\beta,\gamma)\!=\!R_z(\alpha)R_y(\beta)R_z(\gamma)$, which leads to the explicit expression
\begin{equation}
R =
\begin{pmatrix}
c_\alpha c_\gamma-s_\alpha c_\beta s_\gamma
&
-c_\alpha s_\gamma-s_\alpha c_\beta c_\gamma
&
s_\alpha s_\beta
\\
s_\alpha c_\gamma+c_\alpha c_\beta s_\gamma
&
-s_\alpha s_\gamma+c_\alpha c_\beta c_\gamma
&
-c_\alpha s_\beta
\\
s_\beta s_\gamma
&
s_\beta c_\gamma
&
c_\beta
\end{pmatrix},
\end{equation}
with $c_i=\cos(i)$ and $s_i=\sin(i)$.

The 3SM model therefore reduces to the determination of these three Euler angles, which uniquely define the mixing amplitudes between the intrinsic and physical configurations. 
Once the rotation matrix is known, all intrinsic electromagnetic properties and the corresponding effective interaction responsible for the configuration mixing can be reconstructed directly from the experimental observables, as discussed in the following sections.

\subsection{Electromagnetic observables}

Among the possible observables that can be considered to extract the mixing amplitudes, electromagnetic MEs, which can be measured via Coulomb-excitation experiments, represent a particularly straightforward approach due also to the rich set of results. 
Furthermore, such an approach allows the intrinsic shapes of the unperturbed configurations to be determined. 

Based on the linear combination introduced in Eq.~\ref{eq:rotation}, for a given multipolarity the electromagnetic ME between two physical states $| J_i^\pi\rangle$ and $| J_j^\pi\rangle$ can therefore be written as
\begin{equation}
\langle J_i^\pi || \hat{T_\lambda} || J_j^\pi\rangle
=
\sum_{k,l}
a_{ik}
a_{jl}
\langle J_k^\pi || \hat{T_\lambda} || J_l^\pi\rangle,
\label{eq:generalME}
\end{equation}
where $k,l=A,B,C$ label the intrinsic configurations, and $\hat{T_\lambda}$ is the general operator acting between those states. 
In the present work the tensor operator corresponds to the electric quadrupole operator, $\hat{T_\lambda} \equiv \hat{E2}$, since the model is constrained using Coulomb-excitation results: the diagonal MEs determine the spectroscopic quadrupole moments of the physical states, whereas the off-diagonal MEs describe the corresponding transition strengths. 
Together, these observables constitute the experimental input of the 3SM model.

Following the standard assumptions adopted in the traditional 2SM formalism~\cite{Wood1992, heyde2011shape, shapecoexistence2022}, transitions connecting different intrinsic configurations are assumed to be strongly hindered owing to their distinct microscopic structures. 
Under this \textit{no cross-talk} approximation, $\langle J_k^\pi || \hat{T_\lambda} || J_l^\pi\rangle\!=\!\delta_{kl}$ and Eq.~\ref{eq:generalME} can be simplified, becoming
\begin{equation}
\boxed{
M^{(J J')}
=
R^{(J)}
D^{(J J')}
R^{(J')\mathrm T},
}
\label{eq:spectral}
\end{equation}
where $M^{(J J')}_{ij}\!=\!\langle J_i^\pi||E2||{J'}_j^\pi\rangle$, $D^{(J J')}_{kk}\!=\!\langle J_k^\pi||E2||{J'}_k^\pi\rangle$ with $k=A,B,C$. 
The superscripts $(J)$ and $(J')$ emphasize that independent rotation matrices describe the mixing within the two spin-parity subspaces.
This equation constitutes the central relation adopted to solve the 3SM model: the experimentally determined electromagnetic MEs are related to their intrinsic counterparts through the orthogonal transformations defined by the mixing amplitudes. 

It is important to emphasize that the \textit{no cross-talk} approximation is introduced solely to reduce the number of unknown intrinsic MEs and thereby obtain a tractable solution of the system. 
The formal structure of the model, based on orthogonal transformations belonging to the $SO(3)$ group, remains completely general and does not rely on this assumption. 
In principle, the formalism could be straightforwardly generalized to include finite cross-configuration MEs whenever supported by sufficiently complete experimental information.

\subsection{Effective Hamiltonian}

The determination of the rotation matrix provides direct access to the intrinsic wave functions and electromagnetic properties of the coexisting configurations. 
However, an equally important objective is to characterize the interaction responsible for the observed configuration mixing. 
This can be achieved by introducing an effective Hamiltonian defined in the intrinsic basis, whose diagonalization reproduces the experimental excitation energies and wave functions.

Within the intrinsic basis introduced in Eq.~\ref{eq:rotation}, the effective Hamiltonian can be written as
\begin{equation}
H=
\begin{pmatrix}
E_A & V_{AB} & V_{AC}\\
V_{AB} & E_B & V_{BC}\\
V_{AC} & V_{BC} & E_C
\end{pmatrix},
\label{eq:Hamiltonian}
\end{equation}
where $E_A$, $E_B$, and $E_C$ denote the energies of the unperturbed states, while the off-diagonal terms $V_{ij}$ represent the interaction strengths responsible for their mixing. 
The physical excitation energies $E_{1,2,3}$ correspond to the eigenvalues of Eq.~(\ref{eq:Hamiltonian}), so the Hamiltonian satisfies $R^{\mathrm T}HR=\mathrm{diag}(E_1,E_2,E_3)$. 
Then, since $R^{-1}=R^{\mathrm T}$, the matrix elements of the Hamiltonian in Eq.~\ref{eq:Hamiltonian} can therefore be reconstructed from the excitation energy of the physical states, yielding
\begin{equation}
\boxed{
H
=
R
\,
\mathrm{diag}(E_1,E_2,E_3)
\,
R^{\mathrm T},
}
\label{eq:Hreconstruction}
\end{equation}
which provides the complete reconstruction of the effective Hamiltonian directly from experimental observables.


The resulting interaction matrix provides a compact phenomenological description of the configuration mixing, establishing a direct connection between experimentally measured spectroscopic observables and the underlying nuclear interaction responsible for the coexistence of multiple configurations. 
Although the present Hamiltonian is not intended as a microscopic interaction, it offers an intuitive representation of the relative energies of the intrinsic configurations and of the coupling strengths that generate the observed physical spectrum.

\section{Application to the semi-magic nucleus $^{116}$Sn}

The 3SM model was applied to the low-lying $0^+$ and $2^+$ states of $^{116}$Sn using the electromagnetic MEs recently determined through a comprehensive multi-step Coulomb-excitation measurement~\cite{Siciliano116SnArxiv}. 
The present application focuses on the $E2$ operator because of the extensive set of signed MEs provided by this measurement. 
Nevertheless, the 3SM formalism is not restricted to the sole quadrupole observables and can, in principle, be applied to any operator connecting the same physical and intrinsic spaces (see Refs.~\cite{PhysRevC.94.051303, Macchiavelli2017Mg} as examples). 

All experimentally determined $E2$ MEs were fitted simultaneously through the system of equations defined by Eq.~\ref{eq:generalME}. 
Within a common solution, the fit provides the intrinsic $E2$ MEs together with the rotation matrices $R^{(0)}$ and $R^{(2)}$, which describe the configuration mixing
among the physical $0^+$ and $2^+$ states, respectively.
As expressed in Eq.~\ref{eq:spectral}, under the \textit{no cross-talk} approximation the different classes of measured $E2$ MEs constrain complementary properties of the intrinsic configurations: the diagonal and transitional MEs connecting the physical $2^+$ states are principally related to the spectroscopic quadrupole moments of the intrinsic $2^+$ configurations; the $J^\pi\to(J-2)^\pi$ transitions connecting the physical states, on the other hand, principally provide information on the corresponding in-band transition strengths between intrinsic states.

The resulting set of intrinsic $E2$ MEs can then be combined through the quadrupole sum rules~\cite{kumar1972,gosia1986} to characterize the quadrupole deformation associated with each intrinsic configuration. 
In this way, the 3SM model provides not only the mixing amplitudes of the physical states, but also a direct phenomenological reconstruction of the electromagnetic properties of the underlying configurations. 
In this regard, the analysis identifies three markedly different structures: a spherical configuration, a weakly oblate configuration with $\beta_2\!\approx\!0.14$, and a strongly deformed triaxial configuration with $(\beta_2,\gamma)\!\approx\!(0.46,42^\circ)$. 

\begin{figure}[h!]
    \centering
    \includegraphics[width=0.47\textwidth]{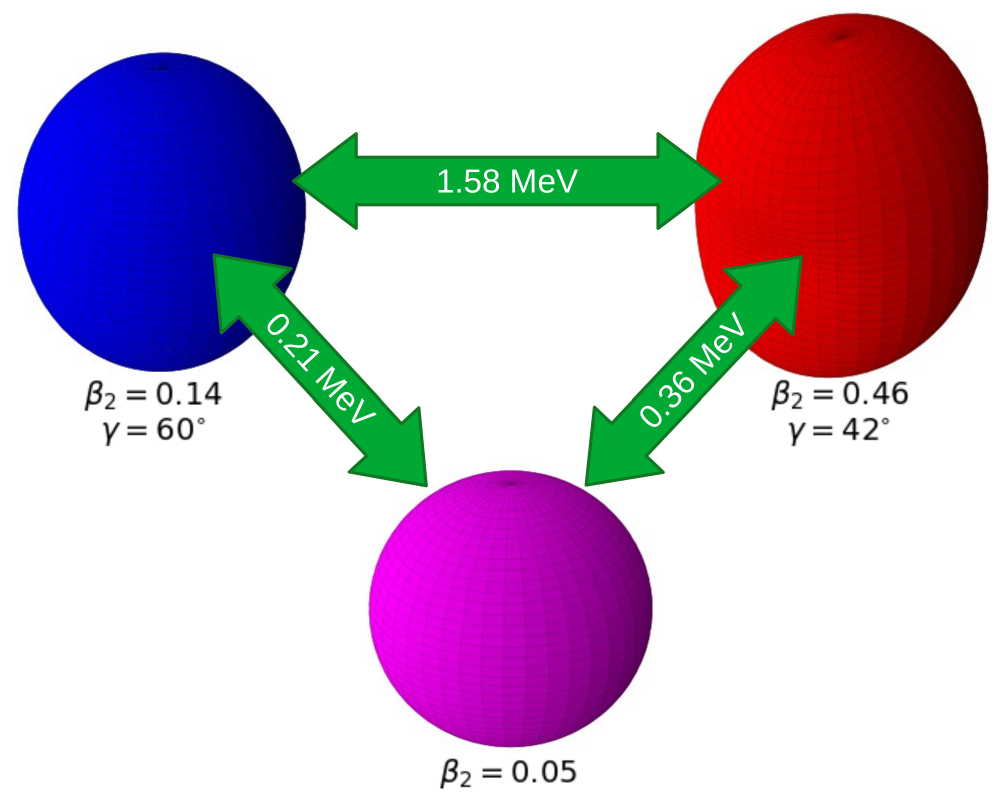}
    \vspace{-3mm}
    \caption{\label{fig:unperturbed}Intrinsic quadrupole shapes of the three configurations extracted for $^{116}$Sn within the 3SM framework. The green arrows indicate the interaction strengths $V_{kl}$ responsible for the mixing.}
\end{figure}

At the same time, the extracted $R^{(0)}$ matrix reveals substantial configuration mixing among the physical $0^+$ states, with no state containing a single intrinsic component exceeding $80\%$. 
The fragmentation is particularly pronounced for the ground states, whose dominant component account for less than $60\%$ of its wave function. 
Such a result further illustrates the unconventional structure of the semi-magic Sn isotopes discussed in Ref.~\cite{Siciliano116SnArxiv}, while a more detailed interpretation of the configuration mixing is deferred to future work. 
In contrast, the $R^{(2)}$ matrix indicates that the physical $2^+$ states are considerably
purer, with dominant configuration amplitudes of about $90\%$. 
This result, obtained within the present simple phenomenological framework, is consistent
with the IBM-2 calculations reported in Ref.~\cite{petrache2019}.

Finally, combining the experimentally known excitation energies of the physical $0^+$ states with the extracted $R^{(0)}$ matrix allows the effective Hamiltonian to be reconstructed through Eq.~\ref{eq:Hreconstruction}. 
Its diagonal elements provide the unperturbed energies of the three intrinsic configurations, while the off-diagonal elements $V_{AB}$, $V_{AC}$, and $V_{BC}$ quantify the corresponding interaction strengths responsible for their mixing. 
The resulting intrinsic shapes and interaction strengths are summarized graphically in Fig.~\ref{fig:unperturbed}. 
It is worth noting that the extracted interaction between the weakly-oblate and strongly-deformed triaxial configurations is particularly large, with $V_{BC}\!\approx\!1.58$~MeV, compared with the $\sim\!200$--$300$~keV strengths obtained for the remaining couplings. 
Such a strong interaction is qualitatively consistent with the exceptionally large $E0$ strength observed between the corresponding mixed structures, further emphasizing their intimate connection. 
At the same time, the magnitude of $V_{BC}$ suggests that the no-cross-talk approximation adopted in the present analysis may represent an overly restrictive description of these two configurations. 
Allowing finite electromagnetic matrix elements between the weakly-oblate and strongly-deformed triaxial configurations may therefore provide a more realistic description of their mixing, an extension that will be explored in a more complete analysis.

\section{Conclusions} 

A Three-State Mixing model has been introduced as a natural extension of the traditional Two-State Mixing approach to systems in which three configurations coexist and interact. 
By describing the transformation between intrinsic and physical states through $SO(3)$ rotations, the framework establishes a direct connection between experimental observables, configuration-mixing amplitudes, and the intrinsic properties of the underlying structures. 
Once the mixing matrix is determined, the same transformation can be combined with the experimental excitation energies to reconstruct an effective Hamiltonian, providing both the unperturbed energies and the interaction strengths among the three configurations. 

The 3SM model has been applied to the low-lying structure of $^{116}$Sn using the extensive set of electromagnetic MEs recently obtained through Coulomb excitation~\cite{Siciliano116SnArxiv}. 
The analysis reveals three intrinsic configurations characterized by markedly different quadrupole shapes and substantial mixing among the physical $0^+$ states, demonstrating the capability of the framework to disentangle the complex structure associated with multiple-shape coexistence. 
While the present contribution focuses on $E2$ observables and $^{116}$Sn, the formalism is independent of the specific operator and microscopic origin assigned to the intrinsic configurations, providing a general phenomenological framework for investigating three-state configuration mixing in finite quantum systems.

It is worth noting that, by following the same procedure presented in this work, the discussed framework can be generalized to the mixing of an arbitrary number $N$ of configurations by replacing the $SO(3)$ transformation with an $SO(N)$ rotation. 
Such an extension, however, rapidly increases the dimensionality of the problem since an $SO(N)$ transformation alone contains $N(N-1)/2$ independent parameters. 
Consequently, increasingly complete experimental information is required to sufficiently constrain the solution. 
The development and experimental requirements of such an $N$-state extension lie beyond the scope of the present work.

\section*{Acknowledgments}

This manuscript owes much to the collaboration with P.E.~Garrett, M.~Rocchini, and T.R.~Rodr\'iguez. 
The author is supported by the U.S. Department of Energy, Office of Science, Office of Nuclear Physics, under contract number DE-AC02-06CH11357. 

\bibliography{116Sn_Ref}

\newpage
\begin{figure*}[b!]
\includegraphics[width=\textwidth]{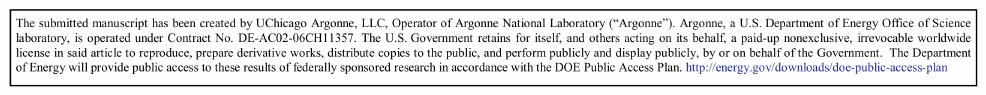}    
\end{figure*}

\end{document}